\documentclass[pre,twocolumn,aps,showpacs]{revtex4}

\usepackage{graphicx}

\begin{document}
\title[Mobile $\pi-$kinks in $0-\pi$ Josephson junction arrays]
{
Mobile $\pi-$kinks and half-integer zero-field-like steps
in highly discrete alternating $0-\pi$ Josephson junction arrays
}
\author{N Lazarides
}
\affiliation{
Department of Physics, University of Crete, 71003, P. O. Box 2208, 
Heraklion,  Greece,\\
and \\
Department of Electrical Engineering, Technological Educational Institute of Crete,
P. O. Box 140, Stavromenos, 71500, Heraklion, Greece 
}
\date{\today}
\begin{abstract}
The dynamics of a one-dimensional, highly discrete, linear array of 
alternating $0-$ and $\pi-$ Josephson junctions is studied numerically,
under constant bias current at zero magnetic field.
The calculated current - voltage characteristics exhibit 
half-integer and integer zero-field-like steps 
for even and odd total number of junctions, respectively.
Inspection of the instantaneous phases reveals that, in the former case,
single $\pi-$kink excitations (discrete semi-fluxons) are supported,
whose propagation in the array gives rise to the $1/2-$step,  
while in the latter case, a pair of  $\pi-$kink -- $\pi-$antikink appears, 
whose propagation gives rise to the $1-$step.
When additional $2\pi-$kinks are inserted in the array, 
they are subjected to fractionalization, transforming themselves into two 
closely spaced $\pi-$kinks.
As they propagate in the array along with the single $\pi-$kink
or the $\pi-$kink - $\pi-$antikink pair, they give rise to 
higher half-integer or integer zero-field-like steps, respectively.
\end{abstract}

\maketitle

\section{Introduction
}
Conventional ($0-$) Josephson junctions (JJs) in the ground state,
have zero $\phi$,
the gauge invariant phase difference of the superconducting order
parameters on each side of the barrier. However, in some cases 
a JJ may have a ground state with  $\phi$ equal to $\pi$ ($\pi-$JJ).
The possibility of a $\pi-$JJ was first pointed out theoretically
in the tunneling through an oxide layer with magnetic 
impurities in SIS junctions, because of negative Josephson coupling
\cite{Bulaevskii}.
Negative coupling also arises in JJs with ferromagnetic intermediate 
layers \cite{Ryazanov,Kontos}. In that case the state of the JJ, 
i.e., a $0-$state or a $\pi-$state depends sensitively 
on the sample design and the temperature \cite{Radovic}.
Another realization of  $\pi-$JJs became possible due to the
$d-$wave symmetry of the order parameter in high$-T_c$ superconductors.
Thus, $\pi-$JJs were demonstrated using twist-tilt,
and $45^o$ tilt $YBCO$ grain boundary JJs \cite{Lombardi,Testa}.
Very recently, a new technique of tailoring the barrier of
superconductor-insulator-ferromagnet-superconductor (low $T_c$)
junctions can be used for accurate control of the critical current 
densities of $\pi-$JJs (as well as $0-$JJs and $0-\pi$ JJs) 
\cite{Weides,Weides1}.
All these junctions are characterized by an intrinsic 
phase shift of $\pi$ in the current-phase relation or, in other words,
an effectivelly negative critical current.

One-dimensional (1D) discrete arrays of parallel biased $0-$JJs 
for which both dimensions are smaller than the Josephson
length $\lambda_{j}$ (i.e., small JJs),
have been studied extensively in recent years \cite{Ustinov,Pedersen}.
Such systems represent an experimental realization of the spatially discrete
sine-Gordon (SG) lattice, whose dynamic behaviour has been modelled 
with the discrete sine-Gordon (DSG) equation.
That equation is formally equivalent to the Frenkel-Kontorova model,
which describes the motion of a chain of interacting particles
subjected to an external on-site sinusoidal potential \cite{Brown}.
Experiments and numerical simulations have shown that,
even with large discreteness, 
the dynamics of a localized kink (fluxon) in the SG lattice exhibit
some features of solitonic nature, close to the properties of the 
continuous SG solitons \cite{Ustinov1}.
However, discreteness also introduces a number of aspects in fluxon 
dynamics which have no counterparts in the continuum system.
The experimental investigation of discrete JJ arrays is a relevant 
issue in superconducting electronics because they are the basis
of the so called phase-mode logic \cite{Nakajima} and of the rapid 
single-flux-quantum (RSFQ) circuits \cite{Buchholz}, for the construction of
superconducting quantum interference filters (SQIFs) \cite{Schultze},
as well as for developing high frequency mixers \cite{Shi}, 
or variable inductors and  tunable filters \cite{Kaplunenko}.

Naturally, the next step in this field was to consider JJ arrays comprising 
$\pi-$JJs.  
The usage of $\pi-$JJs as passive $\pi-$shifters in RSFQ circuits was recently
suggested in reference \cite{Ustinov3}, and realized in \cite{Balashov}.   
Moreover, recent studies (both theoretical and experimental) 
of JJ arrays with $\pi-$JJs have shown some novel 
features arising out of the interplay between $0-$ and $\pi-$JJs
\cite{Ryazanov1,Deleo1,Liu,Tian,Kornev1,Caputo,Rotoli,Kornev2,Kornev,Rotoli1,Li}.
For instance, when a $2\pi$ kink is introduced in a hypothetical parallel array 
with alternating $0-$ and $\pi-$JJs \cite{Chandran},
it will break up into two separate $\pi-$kinks. This effect is 
reffered to as "fractionalization of a flux quantum" and, surprisingly,
it also appears in long continuous JJs with alternating 
$0$ and $\pi$ facets \cite{Susanto}.
In the limit of very short, randomly alternating $0$ and $\pi$ facets,
that problem has been studied in references \cite{Mints,Ilichev}.
Given that $\pi-$JJs can now be formed in a variety of ways, a parallel array
of alternating $0-$ and $\pi-$JJs could, in principle, be made in the lab. 
It is worth noting that parallel JJ arrays consisting of both $0-$ and $\pi-$ 
JJs were also studied as a model of long, faceted, high$-T_c$ bicrystal JJs
with high misorientation angle of the bicrystalline film ($\sim 45^o$)
\cite{Rotoli,Kornev2,Kornev,Lindstrom}.

In the present work, we investigate numerically the fluxon dynamics in a 1D, 
highly discrete array of alternating $0-$ and $\pi-$JJs. 
We are mainly concerned with
linear arrays, with open ended boundary conditions, and different (even or odd)
total number of JJs. In the next section, we describe the array model and 
typical ground states of such arrays, while in section 3 we present numerically
obtained, typical current - voltage characteristics where zero field-like 
(ZF-like) steps 
appear. We finish in section 4 with the conclusions.   
%%%------ figure1-------------
\begin{figure}[!t]
\includegraphics[angle=0, width=.8\linewidth]{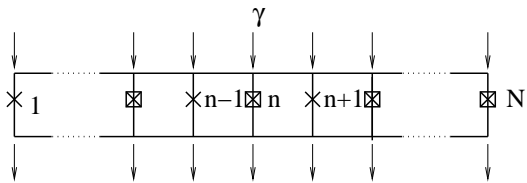}
 \caption{
  Schematic view of a linear array of alternating $0-$ and $\pi-$JJs 
  in parallel.
  Solid lines show the superconducting electrodes,
  while the crosses and the crosses-in-box show the $0-$ and $\pi-$JJs,
  respectively.
}
\end{figure}

\section{Model equations and ground state 
}
Consider a 1D array of alternatingly $0-$ and $\pi-$JJs,
which are connected in parallel via superconducting leads,
as shown schematically in figure 1.
The Hamiltonian function of this system in the simplest approximation,
where all mutual inductances between different cells are neglected, 
is given by 
\begin{eqnarray} 
  \label{1}
    H = e^{-\alpha \, t}  \sum_n  \frac{1}{2} p_n^2 +
    \nonumber \\
     e^{\alpha \, t} \sum_n \left\{ \frac{1}{2} k(\phi_n - \phi_{n-1})^2
     + ( 1 - I_{cn} \, \cos\phi_n ) - \gamma \, \phi_n \right\} ,
\end{eqnarray}
where $\phi_n$ is the phase difference over the $n-$th JJ,
$I_{cn}$ is the critical current of the $n-$th JJ,
$\alpha$ is the dissipation coefficient due to quasiparticles crossing
the barriers, 
$k=1/a^2$ is the coupling strength between nearest neighbouring JJs,
with $a$ being the discreteness parameter,
$\gamma$ is the applied constant (dc) bias current assumed to be injected and 
extracted uniformly (see figure 1), and
\begin{eqnarray}
  \label{2}
  p_n = e^{+\alpha  t} \, \frac{d\phi_n}{dt} .
\end{eqnarray}
is the canonical variable conjugate to $\phi_n$. 
For $\alpha=0$ that variable reduces to the (normalized) 
voltage - phase relation,
$v_n = \frac{d\phi_n}{dt}$, for the $n-$th JJ.
Since $H$ depends explicitly on time $t$, it is obvious that 
it is not a constant of the motion. However, it is interesting that 
the system in consideration belongs in a certain class of systems
having time-dependent Hamiltonians, which may be useful when studying
stability issues of specific solutions.
For $\alpha=0$ and $\gamma=0$ we get the simpler Hamiltonian
\begin{eqnarray} 
  \label{1.1}
    H_s = \sum_n \left\{ 
      \frac{1}{2} \left(\frac{d\phi_n}{dt} \right)^2
     +\frac{1}{2} k(\phi_n - \phi_{n-1})^2 \right\} \nonumber \\
     +\sum_n \left\{ ( 1 - I_{cn} \cos\phi_n ) \right\} ,
\end{eqnarray}
which has been used (with $I_{cn}=const.$)  to describe a harmonic chain 
of particles subjected to an external sinusoidal potential \cite{Brown}.
In this context, the variable $\phi_n$ represents the displacement of a 
particle from its equilibrium position, and the three terms in the earlier
Hamiltonian $H_s$  are interpreted as the kinetic energy of the
particles, the harmonic (elastic) interaction of the nearest neighbouring 
particles
in the chain, and the interaction of the chain with an external on-site
sinusoidal potential, respectively. 
If the ratio $k$ of the coefficients of the elastic coupling energy to 
the on-site potential energy is less than unity, then the system is 
considered to be highly discrete. In respect to kink propagation,
the system is highly discrete if the kink width is of the order of 
lattice spacing $a$.

The $n-$dependent critical current $I_{cn}$ is given in compact form by 
\begin{eqnarray}
  \label{3}
   I_{cn} = I_c [ \xi + \cos((n+1)\pi) ] , 
\end{eqnarray}
where
\begin{eqnarray}
  \label{4}
   I_c = \frac{ I_{c0} + I_{c\pi} }{2}, \qquad 
   \xi= \frac{ I_{c0} - I_{c\pi} }{ I_{c0} + I_{c\pi} }
\end{eqnarray}
with $I_{c0}$ ($I_{c\pi}$) being the critical current of the $0-$ ($\pi-$) JJs.
From equations (\ref{3}) and (\ref{4}) we have that $I_{cn} = I_{c0}$ and $-I_{c\pi}$ 
for odd and even $n$, respectively.
Note that $I_{cn}$ may differ both in sign and magnitude for the $0-$ and $\pi-$JJs.
From the Hamiltonian function given in equation (\ref{1}) we get the equations
\begin{eqnarray}
 \label{5}
  \frac{d^2\phi_n}{d t^2}+\alpha\frac{d\phi_n}{d t} 
      + I_{cn}\sin\phi_n 
      \nonumber \\
   =k(\phi_{n+1}-2\phi_n+\phi_{n-1}) 
   + \gamma .
\end{eqnarray}
%%%----------figure2---------
\begin{figure}[t]
\includegraphics[angle=0, width=.8\linewidth]{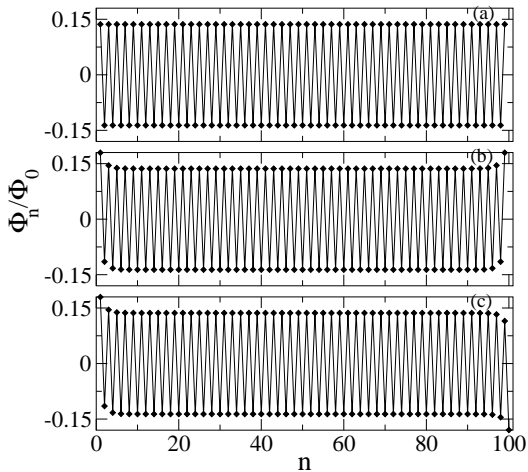}
\caption{
  The normalized flux $2\pi\Phi_n/\Phi_0$ 
  vs plaquette number $n$, for 
  (a) a periodic array ($N=100$);
  (b) an open-ended array with even $N$  ($N=100$);
  (c) an open-ended array with odd $N$ ($N=101$). 
   $I_{c0} = I_{c\pi}$, and $k=0.8$.
}
\end{figure}
For a finite array with $N$ total number of JJs, $n=1,2,...,N$.
The earlier equations may also be obtained by applying Kirchhoff 
laws in the equivalent electrical circuit \cite{Watanabe},
where each single JJ is described by the resistively and capacitively
shunted junction (RCSJ) model. In this model the JJ consists of 
a capacitive branch, a resistive branch, and a superconducting
branch, all connected in parallel.

Equations (\ref{5}) are written in the standard normalization form, 
i.e. $a=D/\lambda_j$  is normalized to $\lambda_j$ 
with $D$ being the spatial separation of neighbouring JJs,
the temporal variable $t$ to the inverse plasma frequency
$\omega_p^{-1}$, and $\gamma=I_c / I_{c0}$ is normalized to $I_{c0}$.
Then, $I_{c0} =1$ in equations (\ref{5}), 
and  $I_{c\pi}$ actually represents the ratio $I_{c\pi} / I_{c0}$.
The voltage $V_n$ across the $n-$th JJ is normalized so that 
$v_n = V_n/V_p$
where $V_p = \Phi_0 \omega_p / (2\pi)$, 
and $\Phi_0 =h/(2e)$ 
is the flux quantum (with $h$ and $e$ being the Planck's constant
and the electron charge, respectively).
Equation (\ref{5}) is analogous to that deduced in reference \cite{Goldobin} 
in the  context of analysis of zig-zag arrays \cite{Smilde}.
Below we deal with arrays both with periodic and open ends, i.e., 
annular and linear finite arrays, respectively. 
Then, in both cases, the governing equations (\ref{5}) remain the same 
for $n=2,...,N-1$.
The same equations are also valid at the end points $n=1$ and $n=N$,
if we define artificial phases $\phi_0$ and $\phi_{N+1}$ such that
\begin{eqnarray}
  \label{66}
     \phi_0 (t) = \phi_N (t) -2\pi M_f \qquad
     \phi_{N+1} (t) = \phi_{1} (t) +2\pi M_f,
\end{eqnarray}
for periodic boundary conditions (e.g., \cite{Watanabe,Pfeiffer}), 
and
\begin{eqnarray}
 \label{6}
   \phi_0 (t) = \phi_1 (t)  \qquad
   \phi_{N+1} (t) = \phi_{N} (t)  ,
\end{eqnarray}
for open-ended boundary conditions. The integer $M_f$ in equation (\ref{66})
is the number of $2\pi$ fluxons trapped in the annular array.
		
Equations (\ref{5}) are integrated with a standard fourth order Runge - Kutta
algorithm with fixed time-stepping (typically $0.01$).
In figure 2 we plot the calculated normalized flux 
$\Phi_n/\Phi_0 = (\phi_n -\phi_{n+1})/(2\pi)$ 
in each elementary cell (plaquette)
as a function of the junction number $n$, for three different arrays in their 
ground state: 
a periodic array (figure 2(a)), and open-ended arrays with even and odd $N$
(figures 2(b) and 2(c), respectively).
In the former case the flux clearly alternates between two values of 
opposite sign and the same magnitude, the latter depending on the coupling 
constant $k$,
indicating that the flux lattice is actually periodic with a period of 
two elementary cells. 
We can thus define a new extented unit cell, with respect to the flux periodicity, 
which consists of two elementary cells. Then, we can evaluate the 
(normalized) average flux
in each of the extended cells, $\Phi_{pm} = (\Phi_{2m-1} + \Phi_{2m})/(2 \Phi_0)$, 
where $m=1,...,N/2$ numbers the extended
cell composed of the ($2m-1$)th and $2m$th elementary cells. 
Clearly, $\Phi_{pm}$ vs $m$ is a constant which equals to zero. 
The same remarks also hold for the open-ended arrays, 
both with even and odd $N$, relatively far from the end JJs.

%%%----------figure3---------
\begin{figure}[t]
\includegraphics[angle=-0, width=.8\linewidth]{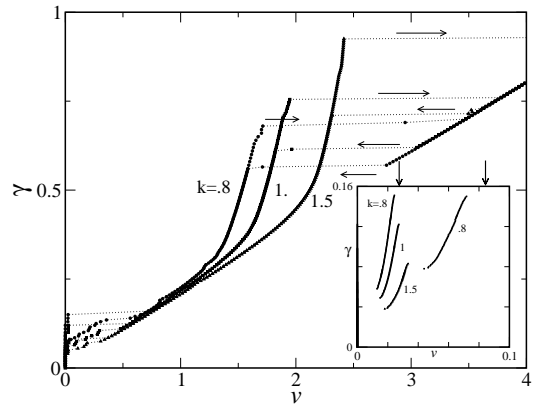}
\caption{
  Current-voltage characteristics for an array with $N$ even ($N=100$),
  $I_{c0} = I_{c\pi}$, $\alpha=0.2$ and $k$ as shown in the figure. 
  The arrows along the dotted lines indicate the direction of the jumps.
  Inset: half-integer ZF-like steps in the lower-voltage part of the curves shown 
  in main figure.
  The vertical arrows indicate the position of $v_m$ for ZFS1/2 and ZFS3/2
  for $k=0.8$ (see text). 
}
\end{figure}

\section{Current - voltage characteristics and zero-field-like steps
}
For obtaining the current-voltage ($\gamma-v$) characteristics for an 
open-ended array,
we initialize the system with $\phi_n =0$ and $\pi$ for $n$ odd and even,
respectively, and $\dot{\phi}_n=0$ for any $n$ at $t=0$ and $\gamma=0$.
(The overdot denotes derivation with respect to the temporal variable.)
Then equations (\ref{5}) are integrated forward in time up to $t_{1}=1000$, 
by which time the system has reached a steady state. 
We then calculate the time averaged voltage $v_n$ across each JJ by averaging
over an additional $t_{2} - t_{1}$ dimensionless time-units, 
i.e., $v_n = \frac{1}{t_2 -t_1} \int_{t_{1}}^{t_{2}} \dot{\phi}_n \, dt$, 
where $t_{2} =2000$.  
The voltage of the array $v$ is obtained by averaging over all JJs,
i.e., $v = \sum_{n=1}^N v_n / N$.
The dc bias $\gamma$ is then increased by $\delta\gamma$ and the calculation 
is repeated,
assuming as initial condition the steady state achieved in the previous calculation. 
We continue this procedure increasing $\gamma$ by $\delta\gamma$ until $\gamma=1.1$, 
then decrease $\gamma$ with same steps back to zero, 
obtaining the $\gamma-v$ characteristics shown in figures 3 and 4 for 
even $N$(=100) and odd $N$(=101), respectively
($\delta\gamma=0.005$ for the curves in the main figures and $0.001$ for the curves 
in the insets). 
Large hysteresis is observed in the high-voltage part ($v \stackrel{>}{\sim} 0.5$)
of all characteristics in both figures, 
even though the damping coefficient is rather high ($\alpha=0.2$).
Small hysteretic effects are also apparent in the low-voltage parts
(not shown for clarity). 
Apparently, the high-voltage characteristics in figures 3 and 4 for same $k$ 
are practically the same.
The solutions on the high voltage branches have been discussed in detail in 
reference 
\cite{Chandran}, and it seems that they are not significantly affected by
the choice of boundary conditions.
Here, we focus on the the low-voltage parts of those characteristics,
which exhibit significant quantitative differences for odd and even $N$.
%%%----------figure4---------
\begin{figure}[t]
\includegraphics[angle=-0, width=.8\linewidth]{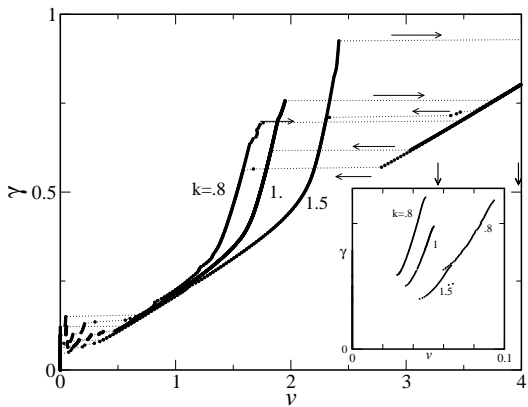}
\caption{
  Current-voltage characteristics for an array with $N$ odd ($N=101$),
  $I_{c0} = I_{c\pi}$, $\alpha=0.2$ and $k$ as shown in the figure. 
  The arrows along the dotted lines indicate the direction of the jumps.
  Inset: integer ZF-like steps in the lower-voltage part of the curves
  shown in main figure.
  The vertical arrows indicate the position of $v_m$ for ZFS1 and ZFS2
  for $k=0.8$ (see text). 
}
\end{figure}
That can be seen by comparing the insets of figures 3 and 4, 
which display a magnified portion of the corresponding 
low-voltage characteristics seen in the main figures.
There we can see the appearence of current steps, 
similar with those observed in $0-$JJ arrays 
(see for example references \cite{Ustinov1,Pfeiffer}).
Those steps are the discrete counterparts of the zero-field steps (ZFSs) appearing in 
the $\gamma-v$ characteristics of continuous, long $0-$JJ, 
due to fluxon resonant motion.
In this case the propagating fluxons are $2\pi-$kinks and their multitudes,
while single $\pi-$kinks are unstable.  
In a continuous, long $0-$JJ, with equivalent normalized length $L\simeq N a$,
the normalized voltage at maximum fluxon velocity $v_m$ of the first ZFS 
would be  $v_m = 2\pi/L \simeq 0.056$ for $k=0.8$.
Thus, the first half-integer ZFS (ZFS1/2) would have $v_m =0.028$,
while the second integer ZFS (ZFS2) and second half-integer ZFS (ZFS3/2) 
would have $v_m =0.112$ and $0.084$, repectively.
The fluxon width decreases (i.e., the fluxon size contracts) with increasing 
velocity.
Thus, in the discrete system, a fluxon cannot reach the maximum velocity
of its continuous counterpart, since its contracted size cannot be smaller than $a$.
In each of the insets in figures 3 and 4 there are two ZFS-like branches 
with $k=0.8$,
with maximum $v$ (defined at the top of the steps) equal to $0.024$, $0.072$, 
and  $0.048$, $0.093$, respectively. 
Thus, the steps in the inset of figure 3 can be identified with the discrete 
counterparts of 
half-integer ZFSs (namely with ZFS$1/2$ and ZFS$3/2$), 
while those in the inset of figure 4 with the discrete counterparts of integer ZFSs 
(ZFS$1$ and ZFS$2$). 

%%%----------figure5-------------------------------------------------
\begin{figure}[t]
\includegraphics[angle=-90, width=.8\linewidth]{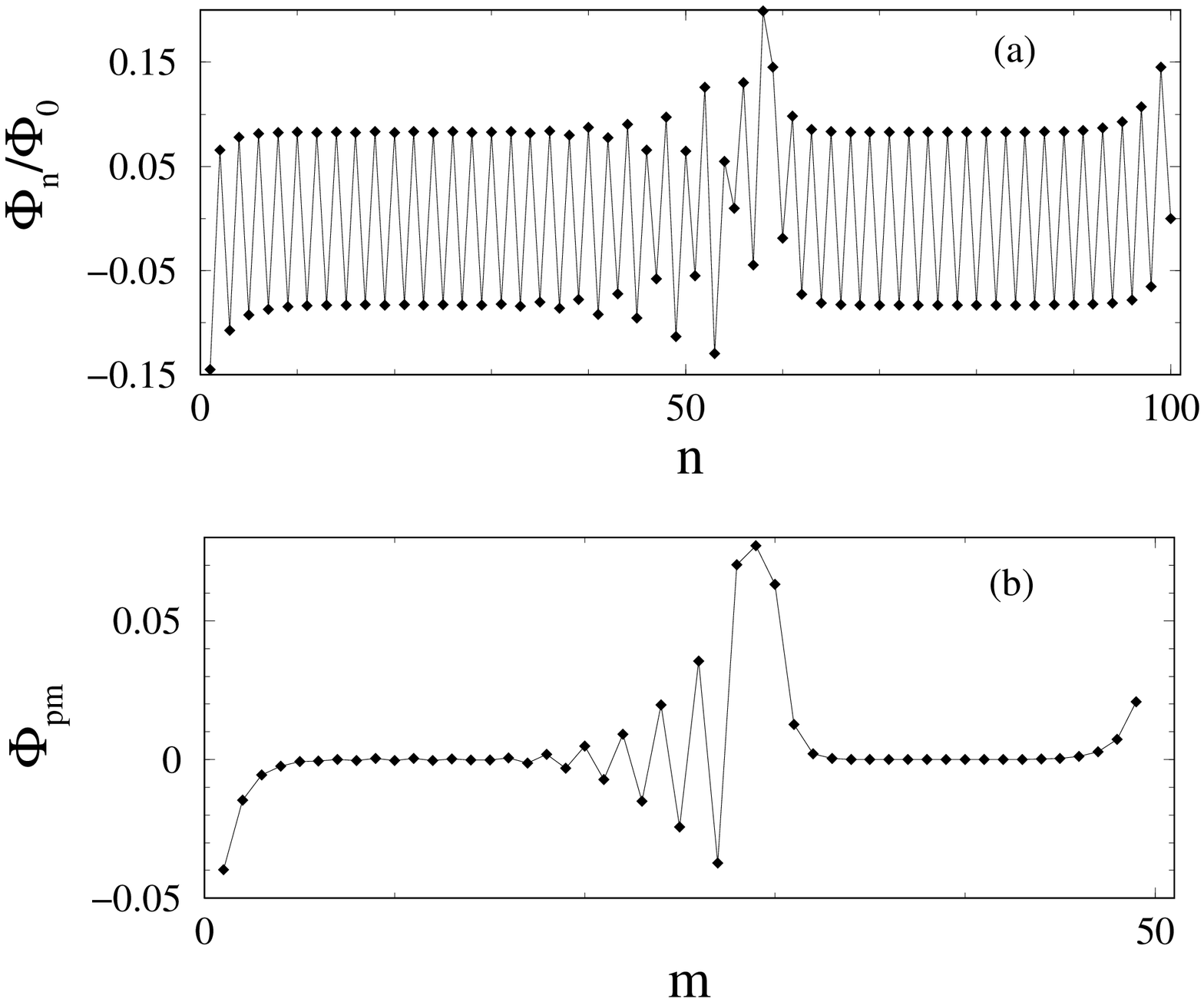}
\caption{
  An instantaneous profile of (a)  $\Phi_n/\Phi_0$ vs  $n$, and  
  (b) $\Phi_{pm}$ vs. extended cell number $m$, for
  $I_{c0} = I_{c\pi}$, $N=100$, $\gamma=0.2$, $k=0.8$, and $\gamma=0.13$
  on the ZFS$1/2-$like branch in the inset of figure 3.
}
\end{figure}
Inspection of the time evolution of the phases (not shown) reveals that the 
ZFS$1/2$-like branch for even $N$(=100) is due to a $\pi-$kink propagating 
in the array,
on a staggered phase background.
In figure 5 we show the instantaneous profile of the normalized flux as a function
of the plaquette number $n$,
and $\Phi_{pm}$ as a function of the number of the extended cell $m$, 
for $\gamma=0.13$.
The antiferromagnetic background in the flux (figure 5(a)) obscures the picture,
but its proper elimination with the doubling of the unit cell along with the
definition of $\Phi_{pm}$ makes things  more clear (figure 5(b)).
Then, in figure 5b we see clearly the discrete $\pi-$kink in a form which is 
the discrete counterpart
of the spatial phase derivative in continuous systems.
Also apparent they are small oscillations behind the $\pi-$kink,
a phenomenon known from $0-$JJ arrays \cite{Ustinov1,Pfeiffer},
where moving $2\pi-$kinks excite linear oscillating modes of the array
(Cherenkov radiation).
Those oscillations, however, get strongly damped and die out fast away from the kink.
In figure 6 the same quantities for an array with the same parameters
but odd $N$(=101) are plotted for $\gamma=0.13$ on the ZFS$1$-like branch.
Here we always have a $\pi-$kink and a $\pi-$antikink pair.
Cherenkov radiation is also apparent here (figure 6(b)) behind both the propagating
$\pi-$kinks.
%%%----------figure6-------------------------------------------------
\begin{figure}[t]
\includegraphics[angle=-90, width=.8\linewidth]{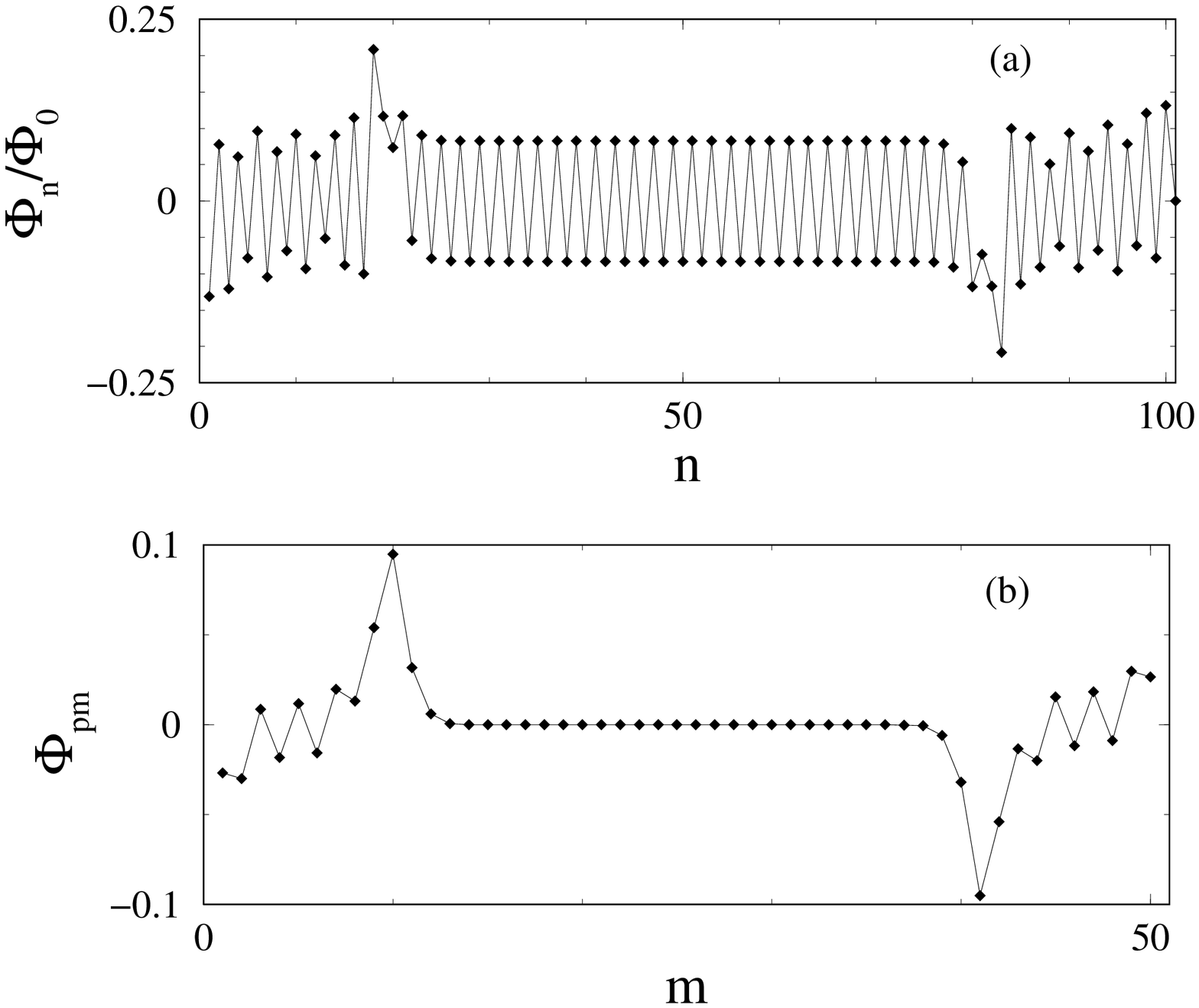}
\caption{
  A snapshot of (a) $\Phi_n/\Phi_0$ vs $n$, and 
  (b) $\Phi_{pm}$ vs. extended cell number $m$, for
  $I_{c0} = I_{c\pi}$, $N=101$, $\gamma=0.2$, $k=0.8$, and $\gamma=0.13$
  on the ZFS$1-$like branch in the inset of figure 4.
}
\end{figure}
Inserting one (or more) $2\pi-$kink(s) in both the even and odd $N$ arrays, 
we get higher order ZFS-like branches. The ZFS$3/2$- and ZFS$2$-like branches
for even and odd $N$, respectively, are shown in the insets of figure 3
and 4, respectively (only for $k=0.8$). 
Those branches were obtained by inserting one $2\pi-$kink in the arrays.
Again, inspection of the phases reveals that the $2\pi-$kink
is fractionalized into two closely spaced but clearly separated $\pi-$kinks
which, however, are not identical.
The snapshot in figure 7 shows the normalized flux as a function of the 
plaquette number $n$,
and $\Phi_{pm}$ as a function of the number of the extended cell $m$, 
for $\gamma=0.13$ on the ZFS$3/2$-like branch in the inset of figure 3
($N=100$, even).
The fractionalized $2\pi-$kink at the right
of figure 7(b), now coexists with the single $\pi-$kink in the array. 
The corresponding figure (figure 8) for the ZFS$2$-like branch in the 
inset of figure 4 ($N=101$, odd), 
shows the fractionalized $2\pi-$kink along with a $\pi-$kink - $\pi-$antikink
pair.

The direction of motion of the fractionalized $2\pi-$kinks depends on their polarity
when they are inserted in the array. These kinks, either the single one or the ones
obtained by fractionalization, behave in many aspects like the usual $2\pi-$kinks.
They get reflected by the ends changing their polarity, and interact
completelly elastically, passing the one through the other without any apparent
shape change. Although not shown in figures 3 and 4, 
the depinning current $\gamma_c$
related to the Peierls-Nabarro potential, which the kink should overcome 
in order to start moving, is practically the same for $N$ even and odd.
It increases with increasing discreteness (decreasing $k$), having the values
$\gamma_c=0.123, ~0.099$ and $0.066$ for $k=0.8, ~1.0$ and $1.5$, respectively.
The fractionalized $2\pi-$kinks shown for a specific illustrative case 
in the figures 7(b) and 8(b), are closely spaced and they move together in 
the same direction without changing their distance of separation. 
Thus, they essentially form a single bound state which resembles the bunched
fluxon state observed in $0-$JJ arrays \cite{Ustinov2}.
Such a state results from strong interactions between kinks, which is 
mediated by their Cherenkov "tails". As a result, the Cherenkov radiation 
behind the bunched state is strongly suppressed, which is exactly what we 
observe in our figures 7(b) and 8(b).

%%%----------figure7---------
\begin{figure}[!t]
\includegraphics[angle=-90, width=.8\linewidth]{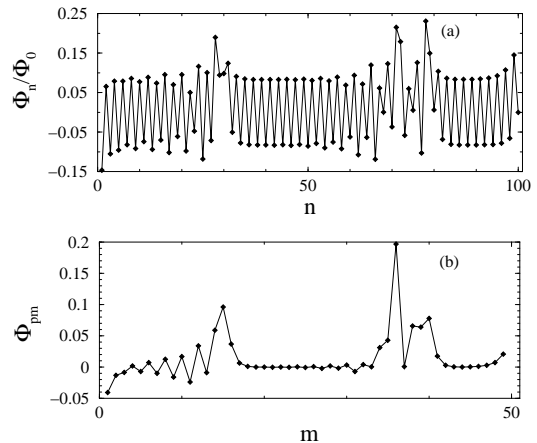}
\caption{
  A snapshot of (a) $\Phi_n/\Phi_0$ vs $n$, and  
  (b) $\Phi_{pm}$ vs. extended cell number $m$, for
  $I_{c0} = I_{c\pi}$, $N=100$, $\gamma=0.2$, $k=0.8$, and $\gamma=0.13$
  on the ZFS$3/2-$like branch in the inset of figure 3.
}
\end{figure}

%%%----------figure8---------
\begin{figure}[h]
\includegraphics[angle=-90, width=.8\linewidth]{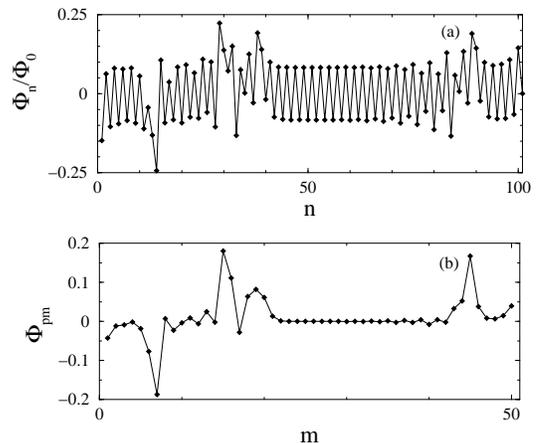}
\caption{
  A snapshot of (a) $\Phi_n/\Phi_0$ vs $n$, and  
  (b) $\Phi_{pm}$ vs. extended cell number $m$, for
  $I_{c0} = I_{c\pi}$, $N=101$, $\gamma=0.2$, $k=0.8$, and $\gamma=0.13$
  on the ZFS$2-$like branch in the inset of figure 4.
}
\end{figure}

In annular $0-\pi$ JJ arrays with both even and odd $N$, 
which satisfy the boundary conditions (\ref{66}), the effect of fractionalization
of $2\pi-$fluxons also appears \cite{Chandran}.
In that case, $2\pi-$fluxons can be inserted in the array either by setting 
$M_f$ to a nonzero integer value or by an appropriate initial condition.
For both even and odd $N$ and a nonzero dc bias $\gamma$ (e.g.,  $\gamma=0.13$), 
there is a number of $\pi-$kinks and
$\pi-$antikinks moving along the array such that the total flux is 
always an integer as required by the boundary conditions.
They appear either as $\pi-$kink - $\pi-$antikink pairs moving in 
opposite directions, or as two bunched $\pi-$kinks looking very much like 
those in figures 7 and 8.
However, we did not find single $\pi-$kinks in this case, so that
half-integer ZF-like branches do not appear.   
Thus, it seems that half-integer ZF-like branches appear only for arrays
with even $N$ satisfying the boundary conditions (\ref{6}),
i.e., for linear arrays. 
This is probably due to the amount of the net total flux in a $0-\pi$
JJ array in its ground state. For arrays satisfying periodic boundary 
conditions without inserted flux, with either odd or even $N$, 
the ground state is always arranged such that the total flux is equal to zero. 
Apparently, this is not always the case for open-ended boundary conditions.
For odd $N$ there is an even number of elementary cells in the array,
so that the contributions from all cells exactly cancel, resulting in
zero total flux. However, for even $N$ there is an odd number of elementary 
cells in the array, and the total flux attains a nonzero value
(approximatelly equal to $0.274$ in figure 2(b)). 
With increasing bias current above the Peirls-Nabarro barrier, 
that ground state becomes unstable against the formation of a $\pi-$kink.
Dynamic simulations for such $0-\pi$ JJ arrays with periodic boundary 
conditions have been performed in 
reference \cite{Chandran}, where it was shown that $2\pi-$kinks 
fractionalize into a pair of $\pi-$kinks in certain regions of 
$k - I_{cp}$ space.
However, ZF-like steps on the current - voltage characteristics are not discussed. 
Although a $0-\pi$ JJ array is still hypothetical, it can be possibly 
realized with the existing technology. 
Moreover,  it has been argued \cite{Kornev2} that bicrystal grain 
boundary JJs with $45^o$ misorientations angle, form intrinsically
an array of alternatingly $0$ and $\pi$ contacts. In this context,
a discrete array of alternating  $0-$ and $\pi-$JJs has been used as 
a model to study the dynamics as well as the  magnetic field dependence 
of such JJs \cite{Kornev2,Kornev}.

\section{Conclusions
}

In conclusion, we caclulated numerically the $\gamma-v$ characteristics for
a 1D discrete, planar array of small, parallel connected alternating $0-$ and 
$\pi-$JJs, which exhibit current steps in the low-voltage part. 
For even $N$, the maximum voltage of the steps (evaluated at the top of each step)
is approximately two times smaller than the maximum voltage of the steps
for odd $N$. 
Comparison with the continuous long JJ, helps in identifying those 
steps as half-integer and integer ZFS-like steps for even and odd $N$, respectively.  
Inspection of the phases and the fluxes reveals that in both cases the system
supports $\pi-$kinks, either a single $\pi-$kink or a $\pi-$kink - $\pi-$antikink 
pair for even and odd $N$, respectively, which can propagate in the array.
Moreover, any $2\pi-$kink inserted in the array is fractionalized into two, 
generally different $\pi-$kinks,
which propagate together in the array along with the single $\pi-$kink
or the $\pi-$kink - $\pi-$antikink pair, for even and odd $N$, respectively, 
giving rise to higher order half-integer and integer ZFS-like branches.
In the continuum, half-integer ZFSs have been observed experimentally
in not very long $0-\pi$ JJs due to the hopping of the semifluxons 
\cite{Goldobin2}.
Similar ZFS-like branches can be found in rather wide parameter range.
The case with $k=0.8$, for which the system is already highly discrete,
was chosen only for better illustration of the half-integer ZFSs.
Those steps  can also be found even for low damping (i.e., $\alpha=0.02$),
as well as for higher damping (i.e., $\alpha=0.42$), which 
is more appropriate for high-temperature superconductor JJ arrays \cite{Smilde}.
Also, they may exist even for slightly different critical currents
(i.e., for $I_{c\pi} =0.9$ for even $N$). 
Dynamic states such as those described above could in principle
be detected by low-temperature
scanning electron microscopy (LTSEM) imaging \cite{Doderer}.

\section*{acknowledgments
}
The author thanks Zoran Radovi\'{c}  for many useful discussions
on $\pi$ junctions.

\end{document}